\documentclass{PoS}
\usepackage{amsmath}

\usepackage{amsfonts}
\usepackage{amssymb}
\usepackage{times}
\usepackage{fontenc}
\usepackage{mathptmx}
\usepackage{graphicx}
\usepackage{subfig}
\usepackage{color}
\usepackage{lineno}
\title{Characterization of the muography background using the Muon Telescope (MuTe)}

\ShortTitle{Characterization of the muography background using MuTe}

\author{\speaker{Jes\'us Pe\~na-Rodr\'iguez}$^{a}$, Luis A. N\'u\~nez$^{a,b}$, and Hern\'an Asorey$^{c,d}$\\
        \llap{$^{a}$} Escuela de F\'isica, Universidad Industrial de Santander, Bucaramanga-Colombia\\
        \llap{$^{b}$} Departamento de F\'{\i}sica, Universidad de Los Andes, M\'erida-Venezuela.\\
        \llap{$^{c}$} Departamento F\'isica M\'edica, Centro At\'omico Bariloche, Comisi\'on Nacional de Energ\'{\i}a At\'omica, Bariloche-Argentina; \\
        \llap{$^{d}$} Instituto de Tecnolog\'{\i}as en Detecci\'on y Astropart\'{\i}culas (ITEDA), Buenos Aires-Argentina.\\
        E-mail: \email{jesus.pena@correo.uis.edu.co},
        \email{lnunez@uis.edu.co}, \email{asoreyh@cab.cnea.gov.ar}
        }

\abstract{In this work, we estimate the background components in muography using the MuTe: a hybrid muon telescope composed of two subdetectors –a scintillator hodoscope and a Water Cherenkov Detector (WCD). The hodoscope records the trajectories of particles crossing the telescope, while the WCD measures their energy loss. The MuTe hodoscope reconstructs 3841 different directions with an angular resolution of 32\,mrad for an inter-panel distance of 2.5\,m. The spatial resolution can reach $\sim$25.6\,m assuming an 800\,m distance to the target. The WCD measures the deposited energy from 50\,MeV to 1.5\,GeV with a resolution of 0.72\,MeV.

MuTe discriminates muography background sources such as: upward coming muons, scattered muons, the soft component of Extensive Air Showers (EAS), and particles arriving simultaneously. They are filtered by using measurements of deposited energy (WCD) and Time-of-Flight. The WCD differentiates single muons, electrons/positrons, and multiparticle events.  On the other hand, the ToF measurements allow us to estimate the muon momentum establishing an energy threshold to decrease the background contribution of scattered muons. Upward coming muons are rejected by means of the particle arrival direction determined by the ToF sign. 

We concluded that near 36$\%$ of the recorded events belong to the electromagnetic component (electrons and positrons), roughly 30.4$\%$ is caused by multiple particle events that arrive with time differences < 100 ns and the last 34$\%$ are caused by muons. The muonic soft component ($<$ 1\,GeV/c) represents 46$\%$ of the single-muon events. The upward going particles add up the 22$\%$ of the total flux crossing the MuTe.

}

\FullConference{40th International Conference on High Energy Physics -ICHEP2020-\\
		July 28th - August 6th, 2020\\
		Prague, virtual conference.}

\begin{document}
\section{Introduction}
Muography is a non-invasive technique for scanning anthropic and geologic structures. Its applications cover several fields: container inspection \cite{Blanpied2015}, archaeological building scanning \cite{Morishima2017, GomezEtal2016}, nuclear plant examination \cite{Fujii2013}, nuclear waste monitoring, underground cavities \cite{Saracino2017}, overburden of railway tunnels \cite{ThompsonEtal2019}, and volcanology \cite{TanakaOlah2019}.

The measured flux gives information about the inner density distribution of the structure, but such a flux is affected by a particle background. This noise emerges as a result of multiple phenomena: upward-coming muons \cite{Marteau2014, Cimmino2017}, soft muons ($<$ 1\,GeV/c) \cite{Nishiyama2014B,Gomez2017}, the electromagnetic component ($e^-$, $e^+$, $\gamma$) of EAS \cite{Lesparre2012} and, multiple particle events.

Several methods have been implemented to reject the background in muography. Time-of-Flight systems reject upward-coming muons, absorbent layer installation stops low energy particles, and extra sensitive layers decrease the probability of multiple particle events \cite{Lesparre2012}.

In this paper, we estimate the contribution of the background components in muography using MuTe. MuTe classifies the background by means of deposited energy and ToF measurements. The WCD discriminates the electromagnetic and multiple particle events while the ToF differentiates the frontal and rear particle flux. The soft muons are rejected by a momentum threshold established using the particle identity and velocity.

\section{The Muon Telescope}

\begin{figure}[!h]
\begin{center}
\includegraphics[width=.8\textwidth]{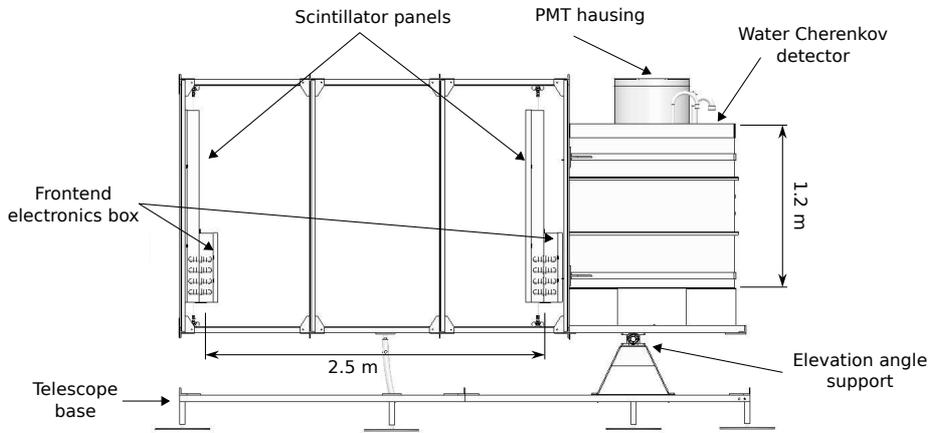}
\caption{Mute mechanical structure. The hodoscope panels are mounted on a sliding rail allowing a separation distance configuration between 0.5\,m to 2.5\,m. The WCD stands near the center of mass of the telescope to facility the elevation procedure. Eight supports affix the MuTe to the ground. }
\label{detector}
\end{center}
\end{figure}

MuTe is composed of a hodoscope and a WCD as shown in Figure \ref{detector}. The hodoscope is made of two scintillator matrices of $30 \times 30$ strips ($120 \textrm{ cm} \times 4 \textrm{ cm} \times 1 \textrm{ cm}$). The plastic scintillator base is polystyrene (Dow Styron $663$) externally coated by TiO$_2$. The dopants ($1\%$  PPO,  $0.03 \%$ POPOP) set an absorption cut off $\sim 40$\,nm and an emission maximum at $420$\,nm. 

Inside each scintillator strip a wavelength shifting (WLS) multi-cladding fiber  (Saint-Gobain BCF-92) transmits the light photons towards a silicon photomultiplier (SiPM, Hamamatsu S13360-1350CS). The SiPM has a photosensitive area of $1.3 \times 1.3 \textrm{ mm}^2$, $667$ pixels, a fill factor of $74\%$, a gain from $10^5$ to $10^6$ and a photon-detection efficiency of $40 \%$ at $450 \textrm{nm}$.

% \begin{figure}[!h]
% \begin{center}
% \includegraphics[width=0.75\textwidth]{Figures/Panel_asemble.jpg}
% \caption{(Left) WLS fiber coupling with the panel aluminum framework. (Right) Top view of the final scintillator panel assembly.}
% \label{panel}
% \end{center}
% \end{figure}

An ASIC MAROC3A amplifies and jointly discriminates the 60 signals from each panel. An FPGA Cyclone 3 sets the ASIC slow control parameters. The data recording is managed by a Raspberry Pi 2 and stored in a central hard disk.

The MuTe-WCD (120\,cm side) has a Tyvek internal coating and an eight-inch photomultiplier tube (PMT Hamamatsu R5912) as the sensitive element. The anode and last dynode  PMT signals are digitized by a 10-bit fast Analog-to-Digital converter with a sampling frequency of 40\,MHz \cite{SofoHaro2016}. The ToF system ($\sim$97\,ps resolution) is based on a Time-to-Digital converter implemented on a Xilinx FPGA Spartan 6 \cite{PeaRodrguez2020}. 
The MuTe data acquisition system is temporally synchronized by GPS. This architecture allows us to correlated the data recorded by the hodoscope and the WCD which operate individually. A GPRS/GSM ITEAD SIM900 module reports daily the telescope state towards a remote server.

\section{Background in muography}

\begin{figure}[!h]
\centering
\includegraphics[scale=0.35]{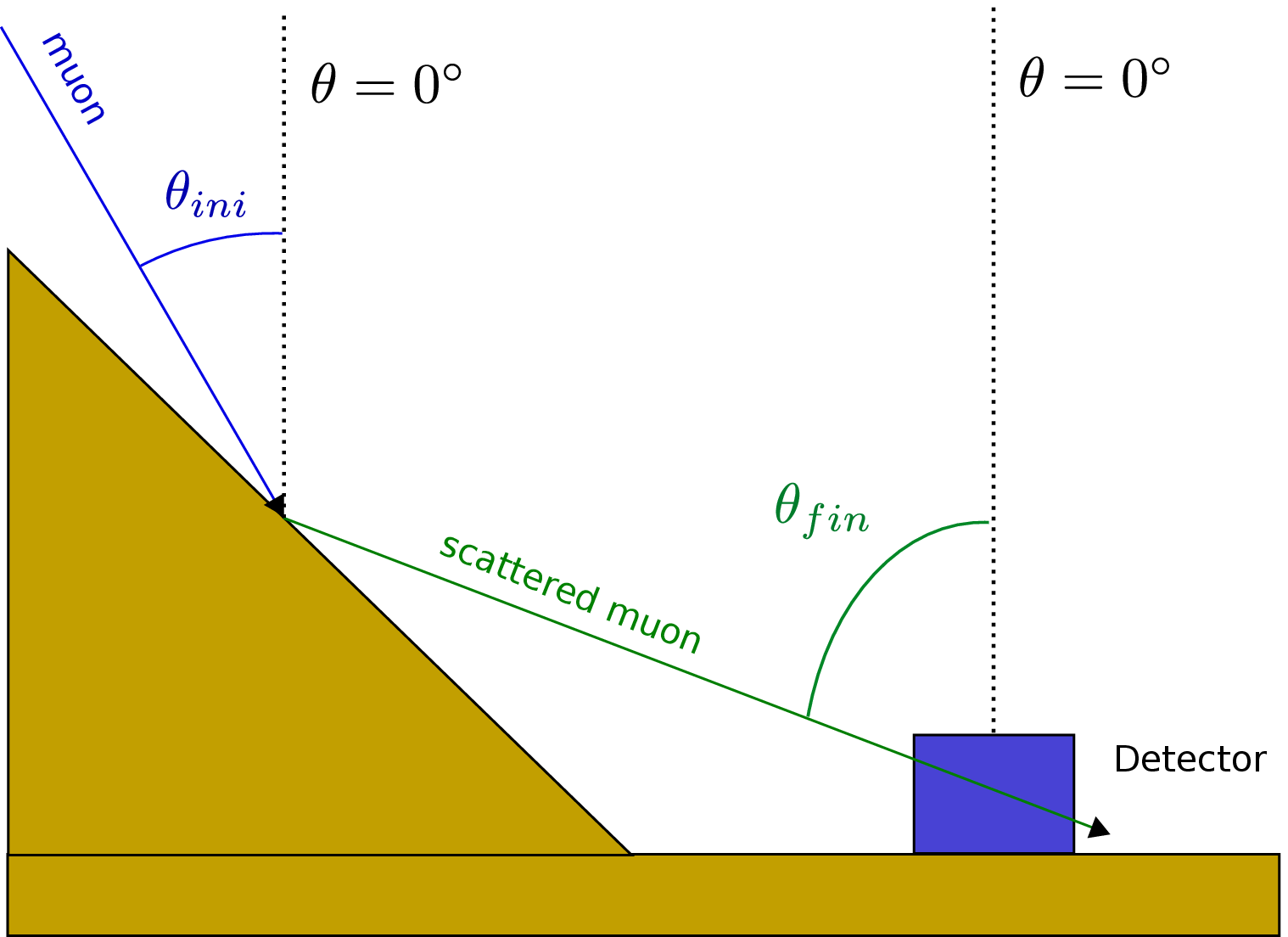} \ \ \
\includegraphics[scale=0.35]{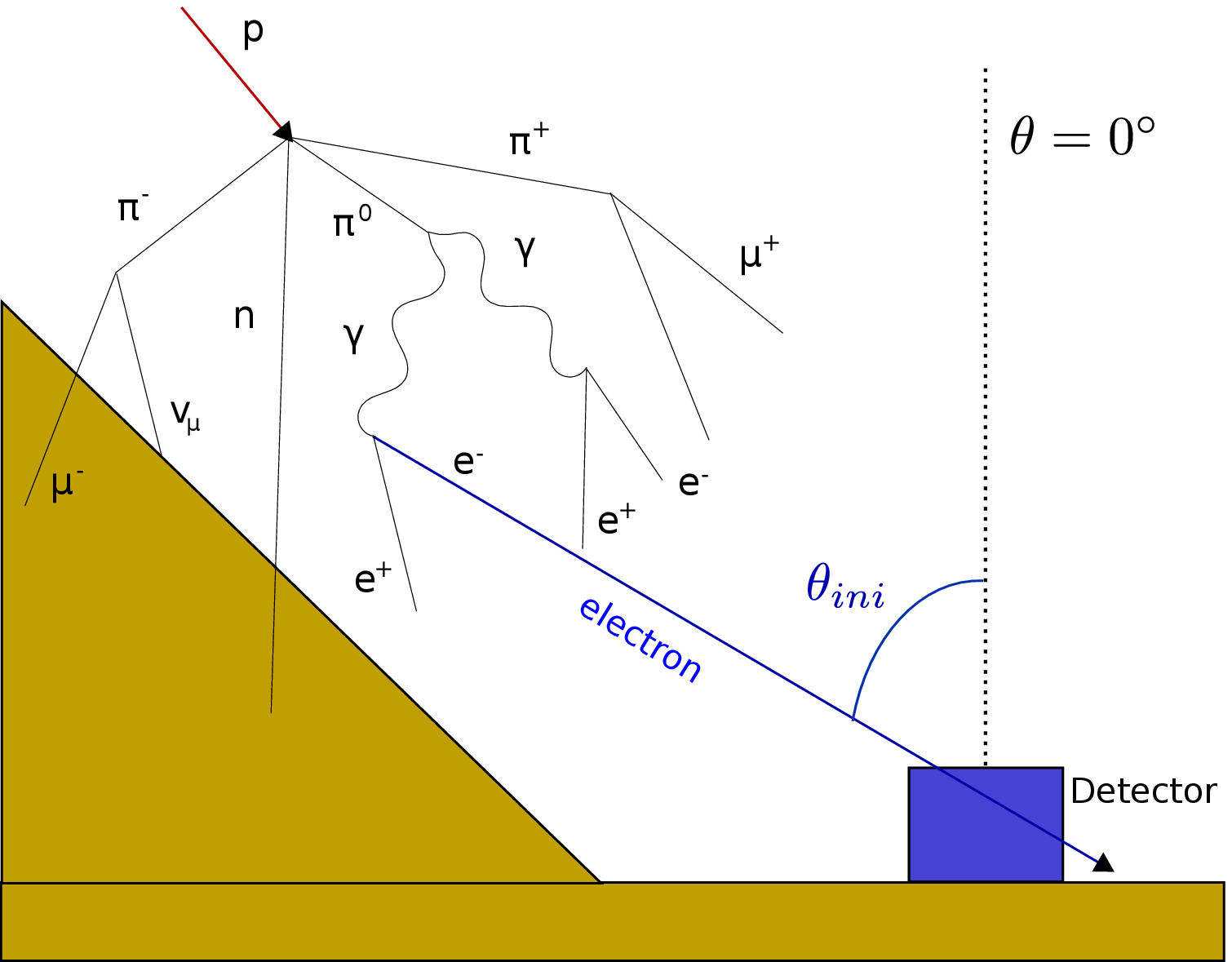}
\includegraphics[scale=0.35]{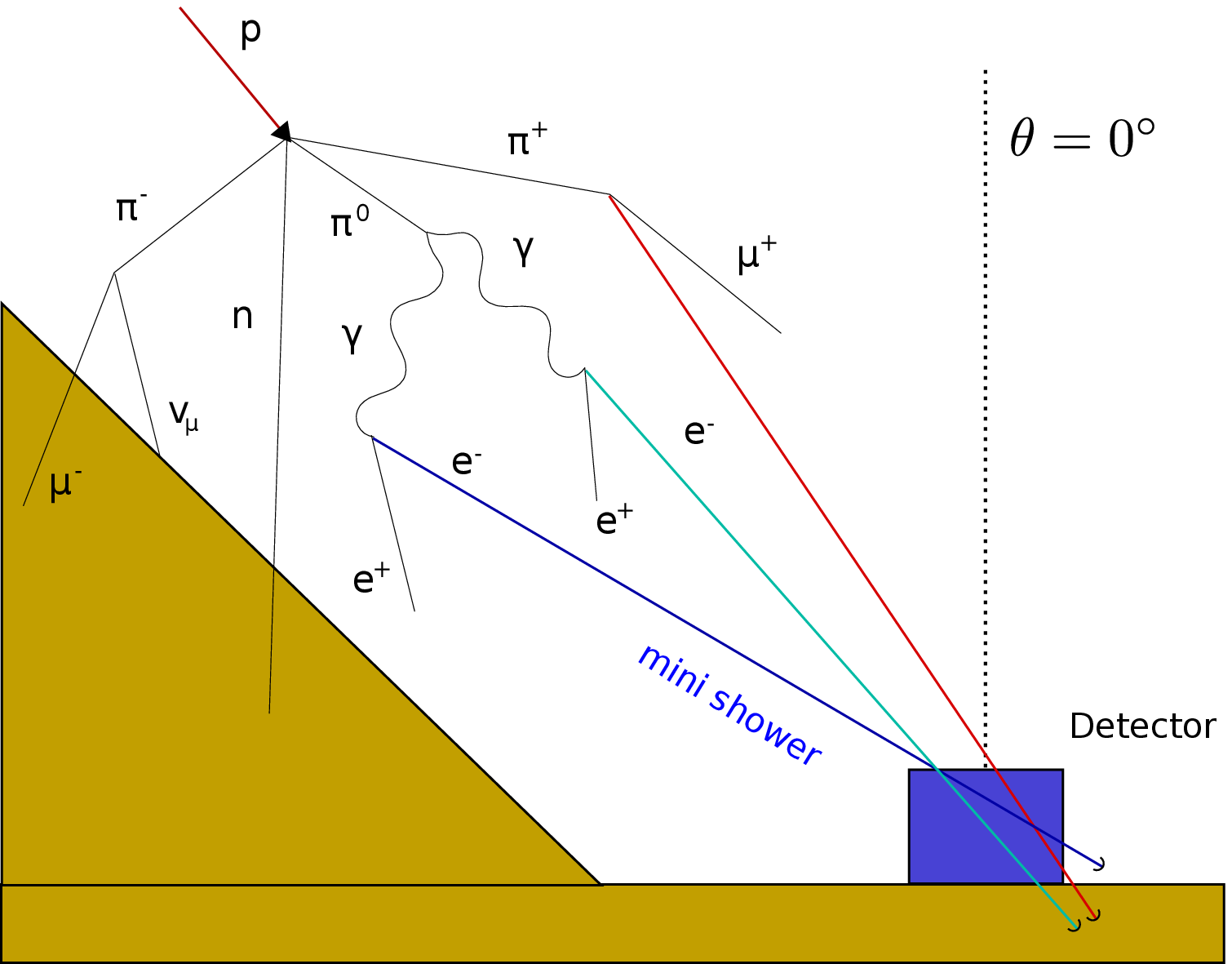} \ \ \
\includegraphics[scale=0.35]{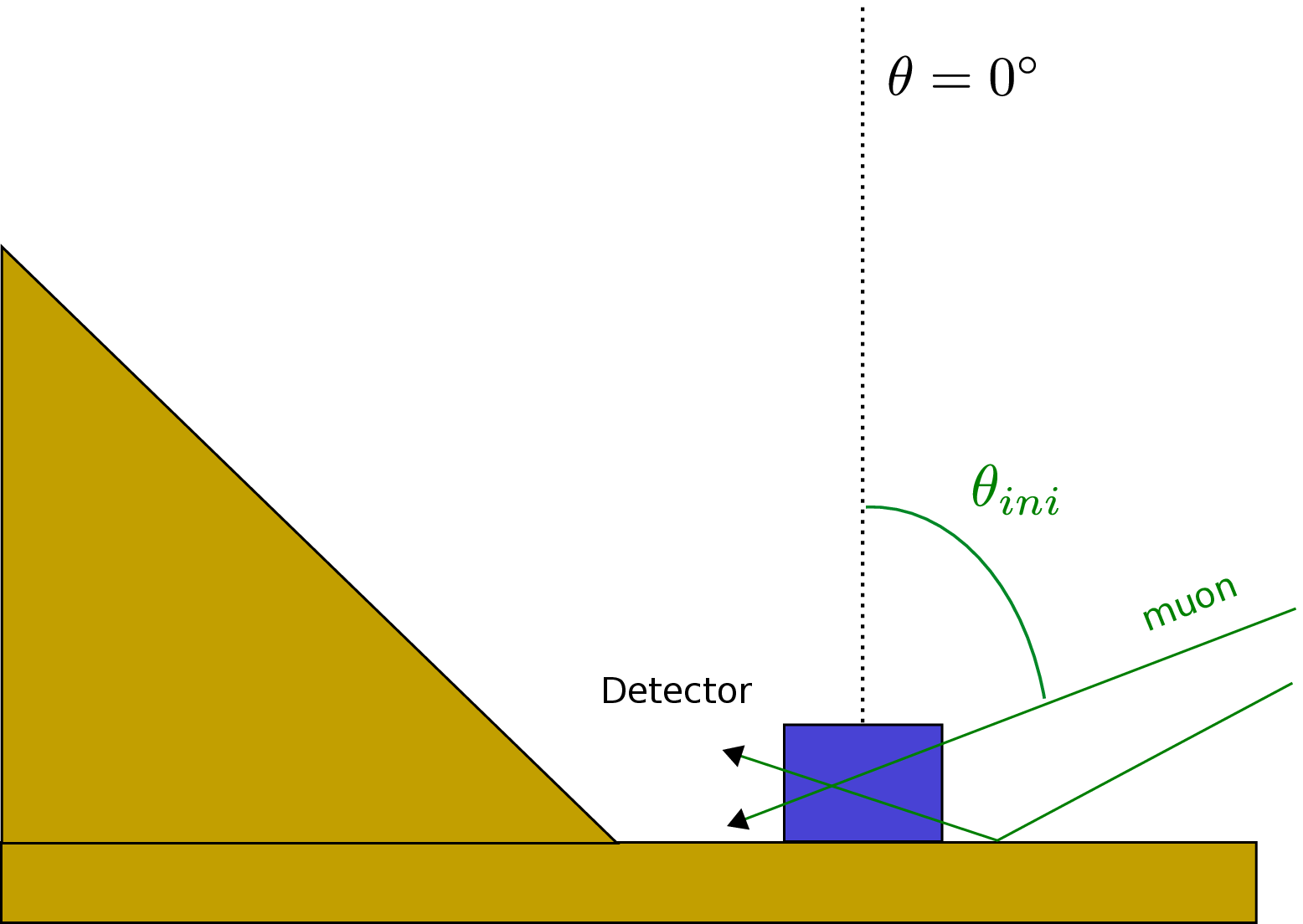}
\caption{Background sources in muography: scattered muons, the electromagnetic component of EAS, mini showers, and upward-coming particles.}
\label{Noise}
\end{figure}

Particles impinging the muon telescope not only come from inside the scanning target. Several underlying phenomena cause an overwhelming particle background that affects the detector signal-to-noise ratio. Muography background is made of: low momentum muons scattered by the target surface, muons entering from the rear side of the detector, charged particles from EAS, and particles arriving simultaneously as shown in Figure \ref{Noise}.

We performed a particle background characterization by means of Monte Carlo simulations by using the \textsl{CORSIKA} code. The results showed that particles reaching the MuTe are mainly positrons/electrons ($\sim$20\,MeV average energy), and muons ($\sim$3\,GeV average energy) \cite{VasquezRamirezEtal2020}.

Low momentum muons suffer trajectory deviation after interacting with the surface of the scanning object due to multiple scattering. If the angular deviation is greater than the telescope resolution, the muogram shall suffer a blurring effect.

The angular variation of the muon trajectory depends on its energy and traversing length inside the target. For instance, muons with energy 3\,GeV crossing 10\,m of standard rock (2.65\,g/cm$^3$) deviate $\sim$35\,mrad \cite{PenaRodrguezPhD2020}.

Scattered and quasi-horizontal ($\theta \geq 75^{\circ}$) muons also arrive from the rear side of the telescope. They have a mean momentum $\sim$10\,GeV/c greater than vertical muons ($\sim$3\,GeV/c). For a telescope elevation angle $<15^{\circ}$ the inverse muon flux can exceed 50$\%$ \cite{Jourde2013}.

Multiple particles simultaneously crossing the telescope can create false-positive events. This combinatorial background is classified as uncorrelated and correlated depending on the origin. Particles coming from independent sources (e.g., different EAS or soil radioactivity) generate the uncorrelated one. The relative arrival time between these particles is in the order of hundreds of microseconds, allowing that ns-resolution detectors can reject them.

Most of the correlated background is composed of muons that originated in the same EAS. For a $<$1\,km distance from the EAS center, muons have a mean arrival delay $<$100\,ns \cite{GarciaGamez2010}. Electron/positrons, generated few radiation lengths ($<$10\,m) close to the telescope, arrive simultaneously within a low relative angle, contributing in this way to the correlated background \cite{Olah2017}.

\section{Background characterization}

\begin{figure}[!h]
\begin{center}
\includegraphics[width=0.55\textwidth]{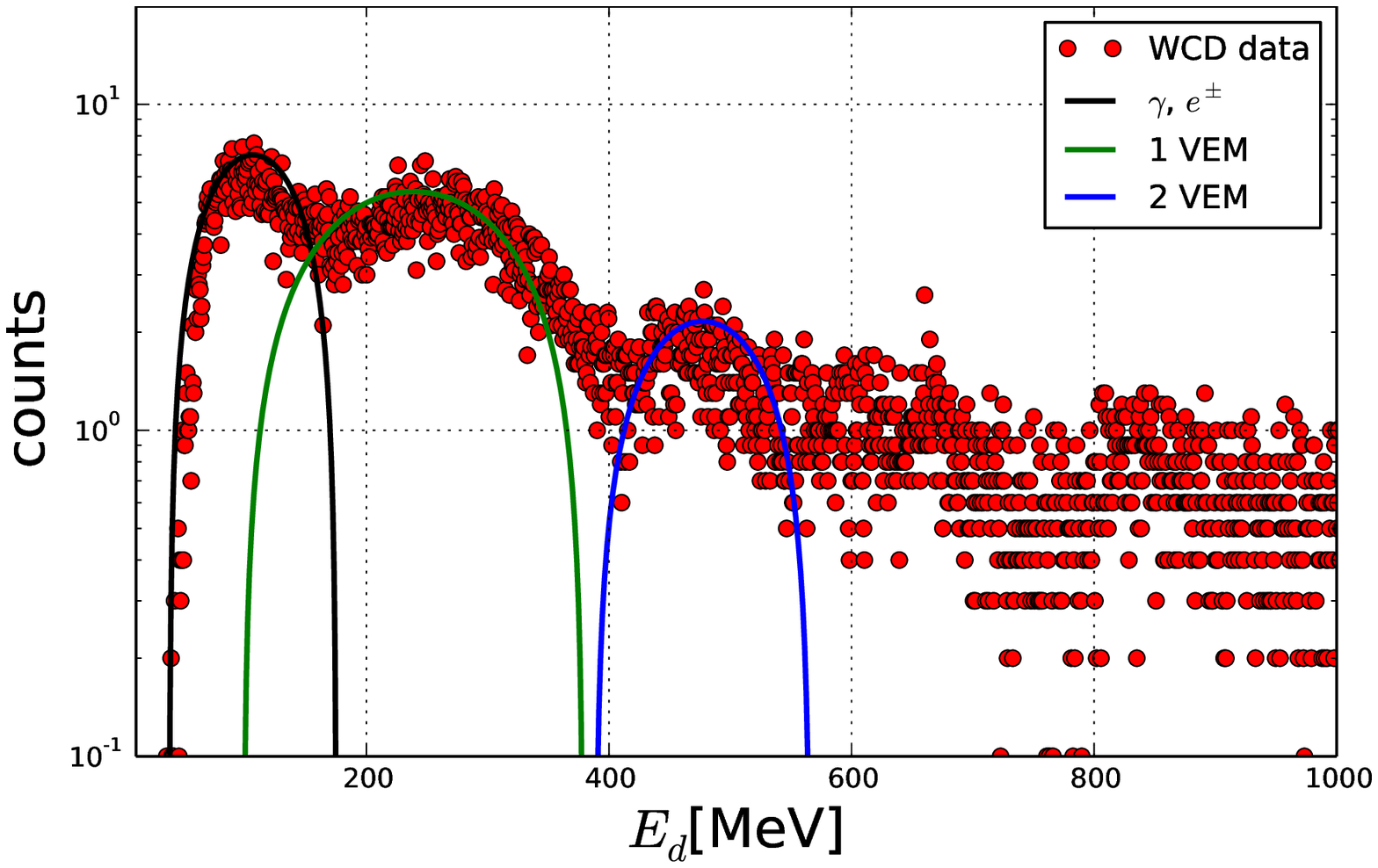}
\includegraphics[width=0.43\textwidth]{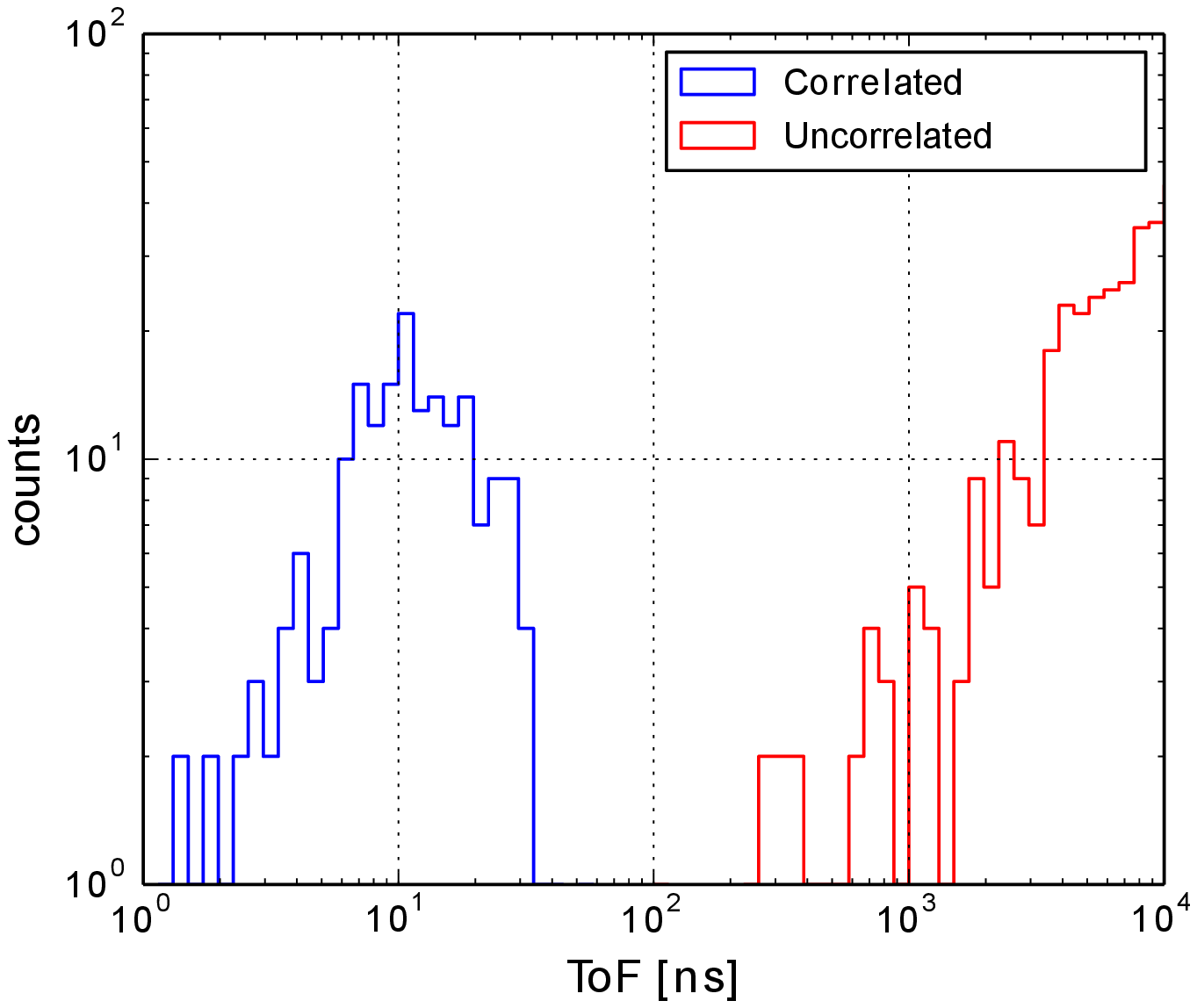}
\caption{(Left) Deposited energy histogram for the events crossing the MuTe during one hour. The black-line hump ($<$180\,MeV) corresponds to the electromagnetic component ($e^{\pm}$, $\gamma$). The green-line hump (180\,MeV$<E_d<$400\,MeV) is the deposited energy of muons. Multiple particle events deposit energy above 400\,MeV. The blue-line hump represents two-muon events. (Right) Time-of-Flight of particles traversing the MuTe hodoscope. Single-particle events and correlated background (blue) have a ToF $<$30\,ns, while the uncorrelated background (red) has a ToF $>$300\,ns.}
\label{Rejection}
\end{center}
\end{figure}

The MuTe recorded data to characterize the particle background at an elevation angle of 15$^{\circ}$ and an inter-panel distance of 2.5\,m during two months. Figure \ref{Rejection}-left displays the energy deposited in the WCD for the particles crossing the hodoscope during one hour. The first hump ($<$180\,MeV) belongs to electromagnetic particles ($e^{\pm}$, $\gamma$) and represents the 36$\%$ of the events.

The muonic component spans from 180\,MeV to 400\,MeV and contains 33$\%$ of the events. The mean of the muonic hump corresponds to the energy released by a vertical muon (VEM) traversing 120\,cm of water --$\sim$240\,MeV. Multiple particle events lose energy above 400\,MeV with a significant contribution at 480\,MeV (2 VEM). These add up the 30.4$\%$ of the events.

The ToF distribution of single-particle events and the correlated background has an upper cutoff at $\sim$30\,ns, and an average value of $\sim$10\,ns as shown Figure \ref{Rejection}-right. The expected ToF of single-particle events spans 2.53\,ns to 20.9\,ns. This estimation was carried out taking into account the scintillator transmission delay ($\sim$77.2\,ps/cm), the hodoscope separation distance (2.5\,m), the scintillator panel dimensions (1.2\,m $\times$ 1.2\,m), and the TDC resolution ($\sim$50\,ps). The uncorrelated background starts at $\sim$300\,ns increasing its occurrence while the time difference grows. The probability that an uncorrelated event occurs below 200\,ns is roughly 0.05$\%$.

\begin{figure}[!h]
\begin{center}
\includegraphics[width=0.9\textwidth]{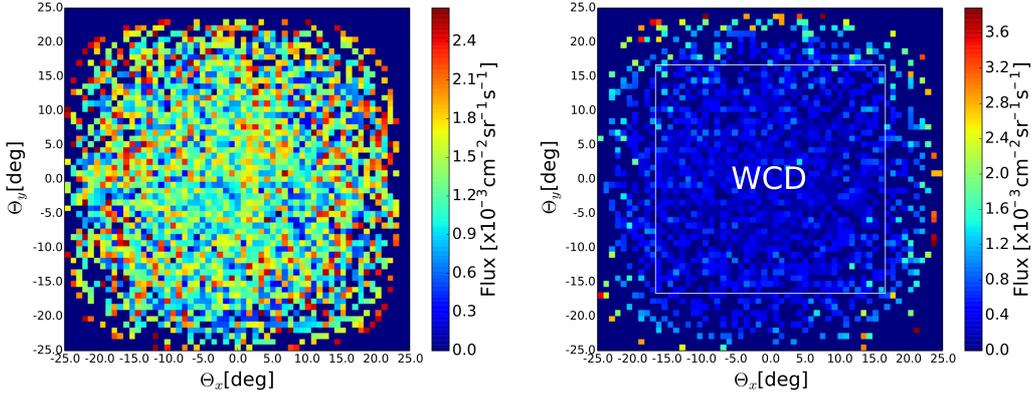}
\caption{Angular distribution of the frontal (left) and inverse (right) particle flux traversing the MuTe hodoscope for an elevation angle of 15$^{\circ}$. The hodoscope axis is located at ($\Theta_x = \Theta_y = 0$\,deg). The frontal flux adds up 78$\%$ of the recorded events reaching up 3.23\,Hz. The WCD water volume (white-square) absorbs the particles with energy $<$ 240\,MeV.}
\label{Backflux}
\end{center}
\end{figure}

Figure \ref{Backflux} compares the frontal (left) and the inverse (right) flux traversing MuTe. The inverse flux represents the 22$\%$ of the particle events impinging the detector. Most of the particles with energy $<$ 240\,MeV are absorbed by the WCD water volume. The inverse flux is higher for inclined tracks than for quasi-perpendicular tracks because of the shadowing effect of the WCD.

\begin{figure}[!ht]
\begin{center}
\includegraphics[width=0.38\textwidth]{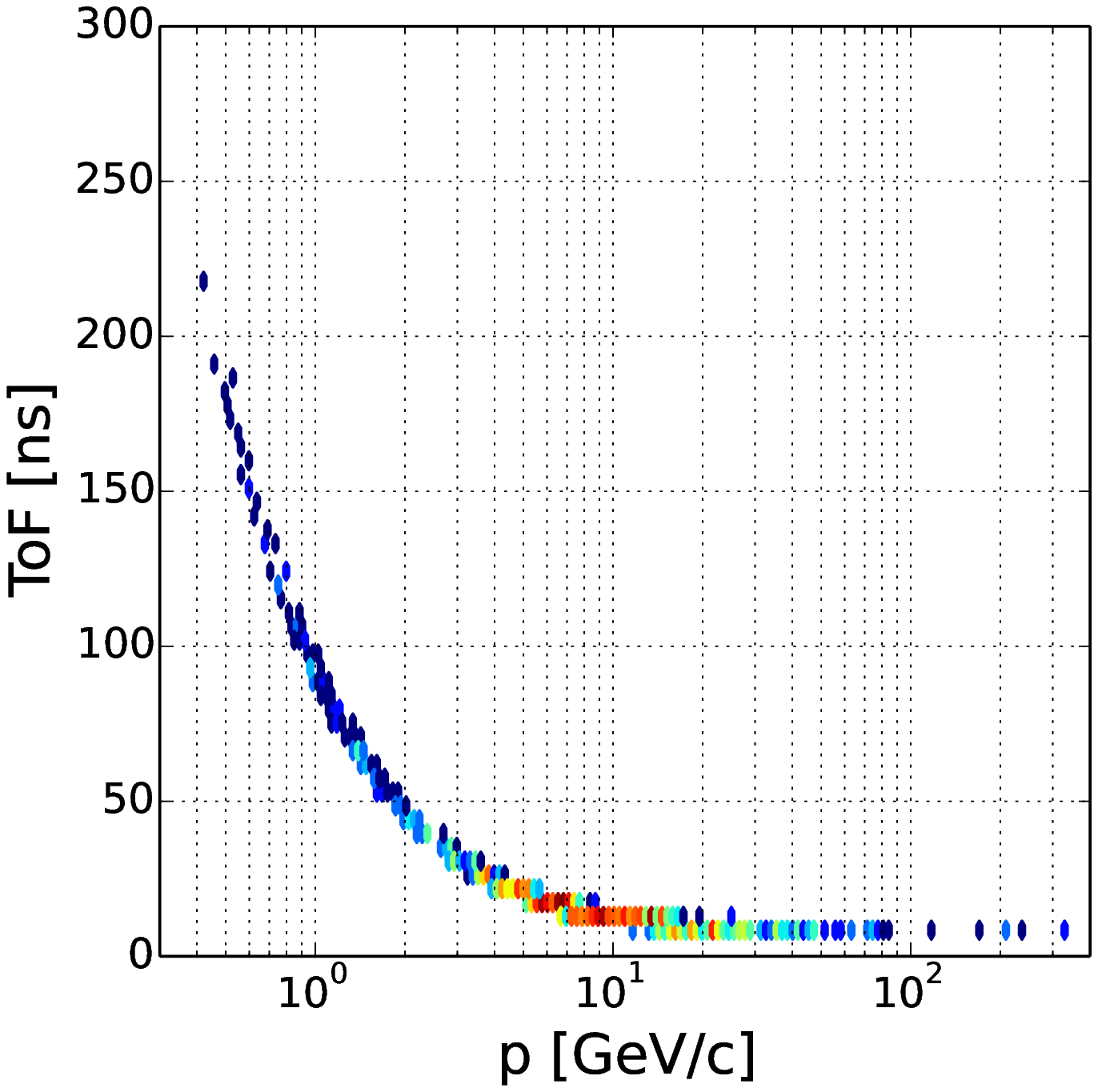}
\includegraphics[width=0.6\textwidth]{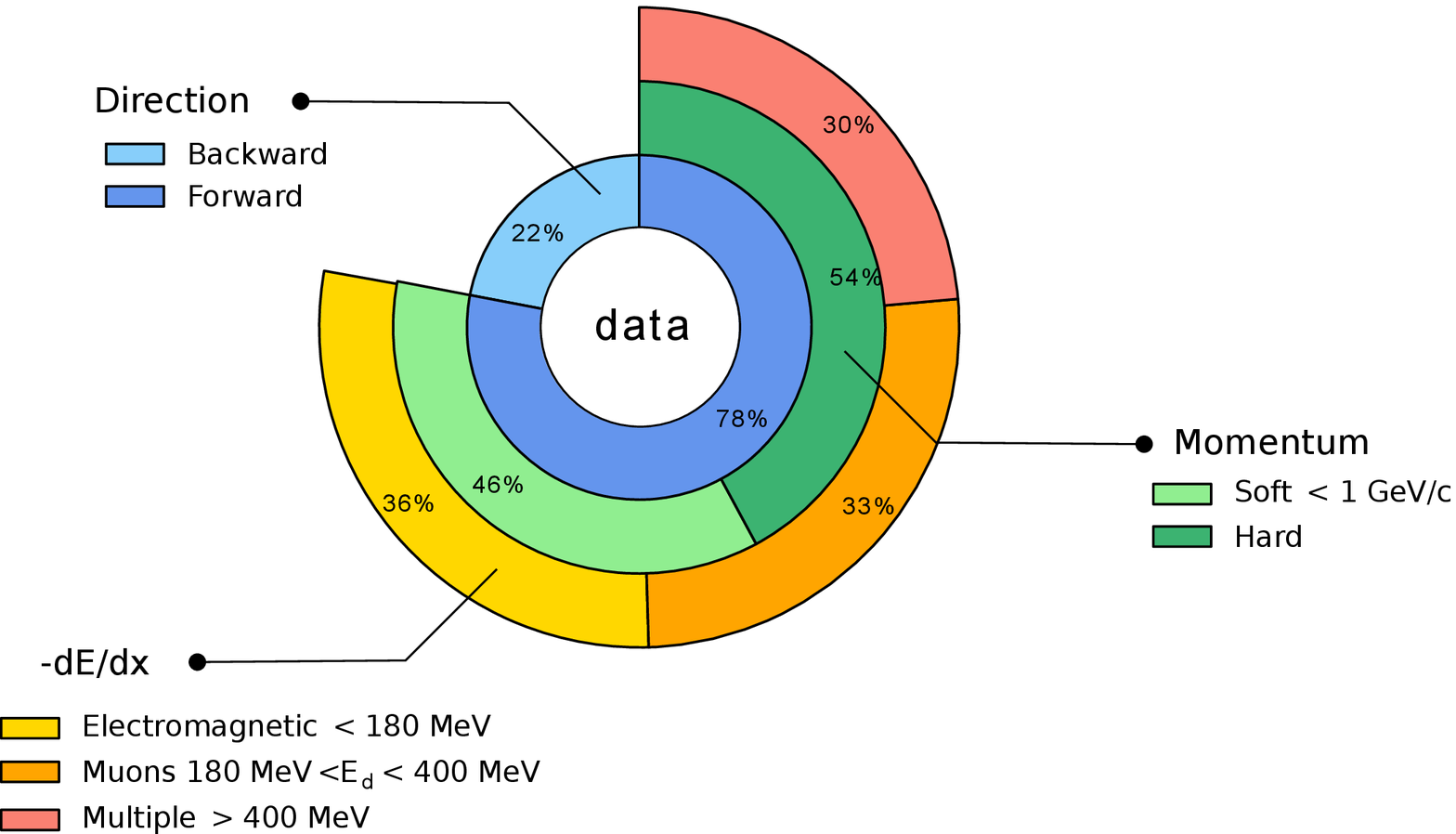}
\caption{(Left) The momentum of particles crossing the MuTe. Events with momentum $>$1\,GeV/c represent 54$\%$ and have an angular deviation $<$ 17\,mrad. (Right) Muography background classification. The frontal flux (78$\%$) is composed of electromagnetic particles (36$\%$), single-muons (33$\%$), and multiple particle events (30$\%$).}
\label{Momentum}
\end{center}
\end{figure}

The incident particle momentum was estimated using the ToF measurements, the trajectory reconstruction, and the particle mass \cite{PeaRodrguez2020}. We set a threshold to reject particles with momentum below 1\,GeV/c as shown in Figure \ref{Momentum}-left. The 54$\%$ of the recorded events have a momentum $>$ 1\,GeV/c and an angular deviation $<$ 17\,mrad. We summarise the results of the muography background characterization in Figure \ref{Momentum}-right.

\section{Conclusions}

We designed and built a muon telescope capable to identify and reject the muography background. The identification methodology of MuTe is based on particle identification techniques -- Time-of-Flight and deposited energy measurements.
We found the background caused by the electromagnetic component of EAS represents $\sim$36$\%$ of the recorded data, while the correlated multiple particle events were $\sim$30.4$\%$. The background due to the inverse flux depends on the elevation angle of the detector: for 15$^{\circ}$ it was $\sim$22$\%$ taking into account the WCD water volume absorbs particles with energy $<$ 240\,MeV. The multiple particle background was caused by two sources: correlated particles with a relative arrival time $<$ 100\,ns, and uncorrelated particles generated by the cosmic ray background or soil radioactivity. As future work in muography, we are developing machine learning techniques to separate automatically the signal from the background.

\acknowledgments
The authors acknowledge the financial support of  Departamento Administrativo de Ciencia, Tecnolog\'ia e Innovaci\'on of Colombia (ColCiencias) under contract FP44842-082-2015 and to the Programa de Cooperaci\'on Nivel II (PCB-II) MINCYT-CONICET-COLCIENCIAS 2015, under project CO/15/02. We are particularly thankful to the Latin American Giant Observatory Collaboration and to Pierre Auger Observatory for their permanent support and inspirations.

%########################################

\bibliographystyle{JHEP}
\bibliography{MuTe.bib}

\end{document}